# Recent global temperature surge amplified by record-low planetary albedo

**Authors:** Helge F. Goessling[1*], Thomas Rackow[2], Thomas Jung[1,3]

**Affiliations:**

[1]Alfred Wegener Institute, Helmholtz Centre for Polar and Marine Research; Bremerhaven, 27570, Germany

[2]European Centre for Medium-Range Weather Forecasts; Bonn, 53175, Germany

[3]Institute of Environmental Physics, University of Bremen; Bremen, 28359, Germany

*Corresponding author. Email: helge.goessling@awi.de

**Abstract:** In 2023, the global mean temperature soared to 1.48K above the pre-industrial level, surpassing the previous record by 0.17K. Previous best-guess estimates of known drivers including anthropogenic warming and the El Niño onset fall short by about 0.2K in explaining the temperature rise. Utilizing satellite and reanalysis data, we identify a record-low planetary albedo as the primary factor bridging this gap. The decline is caused largely by a reduced low-cloud cover in the northern mid-latitudes and tropics, in continuation of a multi-annual trend. Understanding how much of the low-cloud trend is due to internal variability, reduced aerosol concentrations, or a possibly emerging low-cloud feedback will be crucial for assessing the current and expected future warming.

**Main Text:**

After the last major El Niño event in 2015/16 global mean warming was offset by a transition to persistent La Niña conditions in the tropical Pacific (*Iwakiri2023*; Fig. 1a,b). Since March 2023, however, global sea-surface temperatures have broken records (*C3S2024*), well ahead of substantial contributions from the more moderate 2023/24 El Niño (*Schmidt2024*). With the annual global-mean surface temperature 1.48K above the pre-industrial level, in particular the North Atlantic made headlines with the average surface temperatures exceeding previous records by clear margins (*C3S2024*). At the same time, the Antarctic sea-ice extent, after decades of a surprising stability until 2015 (*Rackow2022*), fell below anything that has been previously observed in 2022/23 (*Kuhlbrodt2024,Roach2024*).

Besides the onset of El Niño and the expected long-term warming due to anthropogenic greenhouse gases, there are several factors that may have contributed to the anomalous global-mean temperatures in 2023 (*Schmidt2024*). The 11-year solar cycle is approaching its intensity maximum (*Kopp2016*); the submarine volcano Hunga Tonga—Hunga Ha'apai has released large amounts of water vapor into the stratosphere (*Schoeberl2024*); and new ship fuel regulations, aimed at reducing sulfur emissions, were implemented in three phases, in 2010, 2015 and 2020 (*Hansen2024*). While these regulations may be associated with a spatial pattern that is roughly consistent with the pronounced warming of the traffic-heavy North Atlantic and despite further evidence for recent warming due to reduced aerosols (*Manshausen2022,Hansen2024,Hodnebrog2024*), it has been estimated that the combined global effect of all three factors is below 0.1K and that an unexplained warming of about 0.2K remains (*Schmidt2024*). Based on CERES-EBAF (hereafter CERES) data (*Loeb2018,Minnis2021; Methods*), the recent warming has been linked to an unusually large total top-of-atmosphere (TOA) energy imbalance (EEI; *Kuhlbrodt2024*).

We use CERES satellite and ERA5 reanalysis (*Hersbach2020,Bell2021; Methods*) data to explore the causes of the temperature surge. As synthesized in Fig. 1 and detailed below, we find that:

- The unusually large recent imbalance was mainly driven by a record-low planetary albedo in 2023, continuing a multi-annual trend related to decreasing shortwave reflection by clouds (consistent with *Loeb2024, Song2024*).
- The cloud-related albedo reduction is largely due to a pronounced decline of low-level clouds over the northern mid-latitude and tropical oceans, in particular the Atlantic.
- The increased absorption of shortwave radiation since December 2020 due to reduced albedo can explain about 0.2K of the 2023 temperature anomaly, including about 0.03K from polar regions where declining albedo is dominated by sea-ice and snow retreat.
- Increased incident solar radiation associated with a strong current solar-cycle maximum, captured by CERES but absent in ERA5, has contributed about 0.03K to the 2023 temperature anomaly, whereas El Niño has added about 0.07K.
- Disentangling contributions to the low-cloud trend from internal variability, indirect aerosol effects, and a possibly emerging low-cloud feedback remains challenging.
- Quantifying these contributions will have major implications for the estimated climate sensitivity as well as the current long-term warming level, which may be closer to +1.5K than previously thought.

**Record-high Earth's energy imbalance and planetary albedo**

Earth's energy imbalance (EEI) has been positive for many decades due to the increasing levels of greenhouse gas concentrations from human emissions (*Loeb2018*). According to CERES, the imbalance has been increasing from 2000 onward (*Loeb2021*, *Hansen2024*, *Kuhlbrodt2024*, *Loeb2024*, Fig. 2c), reaching a rate of +0.76Wm$^{-2}$dec$^{-1}$ during the decade prior to 2023 (Tab. 1). However, a record high was reached in 2023 with an anomaly relative to 2001–2022 of +0.97Wm$^{-2}$. ERA5 agrees on the positive sign but exhibits lower values, in particular for the 2023 imbalance (+0.31Wm$^{-2}$; Tab. 1, Fig. 2c).

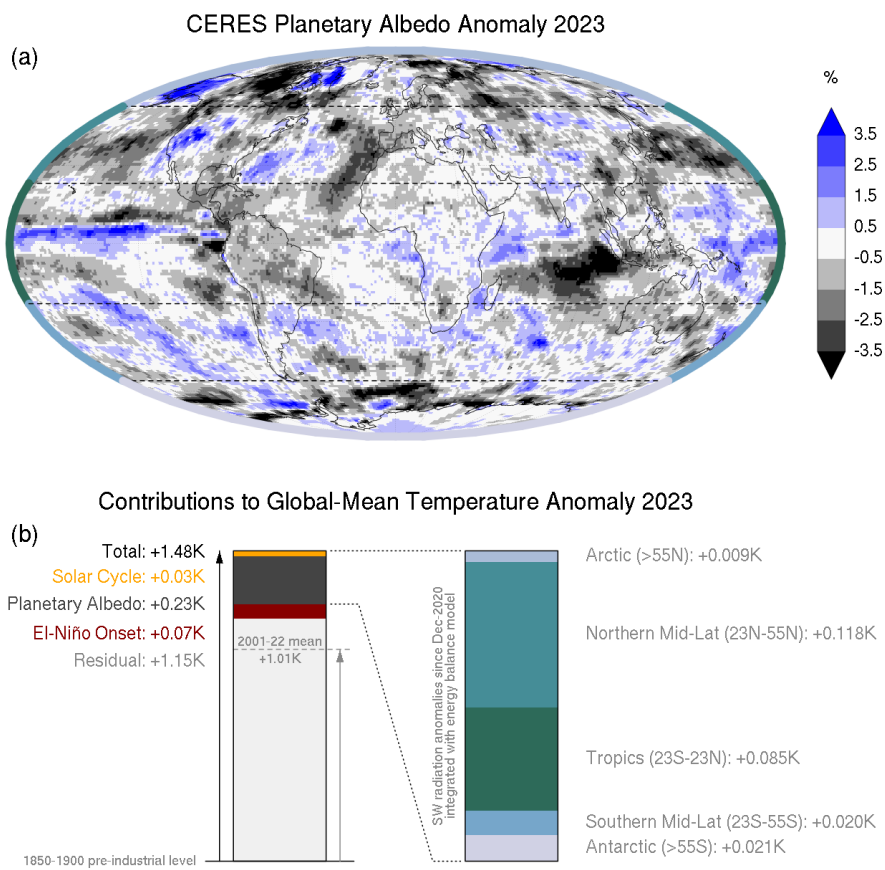

*Fig. 1. Synthesis of contributions to Earth's surface temperature anomaly 2023.* (a) CERES planetary albedo anomaly 2023 relative to 2001–2022 (derived from annual-mean TOA incident solar and upwelling solar radiation); (b) best-guess estimates of contributions to the global-mean surface temperature anomaly 2023 from this study, (left) from the El Niño onset (red) and from absorbed solar radiation (ASR) anomalies since December 2020 due to incident solar radiation anomalies (yellow) and planetary albedo anomalies (dark gray), and (right) the ASR-related contributions further decomposed into contributions from five zonal bands.

The EEI trend and 2023 peak are not associated with decreasing outgoing longwave radiation (OLR), as one would expect from increasing greenhouse-gas concentrations in the absence of shortwave feedbacks. Instead, OLR has been increasing and largely offsetting even stronger absorbed solar radiation (ASR) anomalies (*Loeb2021*, *Loeb2024*, Fig. 2d), consistent with climate models (*Donohoe2014*). The decadal 2013–2022 trend in ASR amounts to +1.10Wm$^{-2}$ in CERES and +0.97Wm$^{-2}$ in ERA5, reaching astonishing +1.82Wm$^{-2}$ in CERES and +1.31Wm$^{-2}$ in ERA5 in 2023 (Tab. 1, Fig. 2c). Variations of incident solar radiation (ISR), including by the 11-year solar cycle, are an order of magnitude smaller (*Kopp2016*, *Schmidt2024*), implying that reduced planetary albedo is the dominant cause (Figs. 1a,S2b, Tab. 1). It is however striking that, according to CERES, ISR attained a positive anomaly in 2023 of +0.28Wm$^{-2}$, well above the previous solar-cycle maximum, whereas ERA5 forcing still assumed a negative anomaly of -0.08Wm$^{-2}$ (Fig. S2f, Tab. 1). Given an absolute planetary albedo of about 29% (Fig. S1i), about 0.20Wm$^{-2}$ of the 2023 ASR and imbalance anomalies in CERES can be explained by the ISR peak, and close to half of the discrepancies between CERES and ERA5.

|  | T<br>*K* | EEI<br>*W/m²* | ASR<br>*W/m²* | CREtc<br>*W/m²* | TC<br>*%* | LC<br>*%* | PA<br>*%* | ISR<br>*W/m²* |
|---|---|---|---|---|---|---|---|---|
| CERES Trend 2013–2022 | – | +0.76 | +1.10 | +0.47 | -0.37 | – | -0.33 | -0.01 |
| ERA5 Trend 2013–2022 | +0.24 | +0.18 | +0.97 | +1.24 | -1.16 | -1.27 | -0.35 | -0.34 |
| CERES Anomaly 2023 | –<br>(+0.22)* | +0.97<br>(-0.44) | +1.82<br>(+0) | +0.58 | -0.34 | – | -0.48 | +0.28 |
| ERA5 Anomaly 2023 | +0.47<br>(+0.31) | +0.31<br>(-0.74) | +1.33<br>(+0) | +1.21 | -0.89 | -1.51 | -0.41 | -0.08 |

**Table 1. Decadal trends for 2013–2022 and anomalies for 2023 of selected global-mean quantities related to Earth's temperature, clouds and energy budget.** *Trends are per decade and 2023 anomalies are relative to 2001–2022. T = surface (skin) temperature; EEI = Earth's TOA total energy imbalance; ASR = TOA net solar radiation (= absorbed solar radiation); CREtc = TOA solar cloud radiative effect inferred from total cloud cover anomalies; TC = total cloud cover fraction; LC = low-level cloud cover fraction; PA = planetary albedo (derived from global-mean TOA solar downwelling and upwelling radiation); ISR = TOA incident solar radiation. Cyan numbers correspond to counterfactuals based on a 2-layer energy budget model (EBM) where ASR anomalies are assumed to be zero from the beginning of December 2020 onward (as in Figs. 1 and S2). (\*derived by combination of the CERES EBM result with the ERA5 temperature anomaly; Methods)*

Long-term trends in ERA5 can be spurious also due to observing-system changes (*Hersbach2020*, *Bell2021*). ERA5 however suggests that planetary albedo was relatively low around the 1940's and 50's (Fig. S3i), before industrial aerosol precursor emissions led to global dimming until the 1980's (*He2018*). The strongest planetary albedo excursions were high-albedo episodes caused by volcanic eruptions, with annual ASR anomalies reaching -3Wm$^{-2}$ in 1992, after the Mount Pinatubo eruption (Fig. S3d). Negative albedo anomalies below the 1950's minimum were however absent, suggesting that the 2023 planetary albedo may have been the lowest since at least 1940.

**Most pronounced changes in the northern hemisphere and tropics**

Positive ASR anomalies in 2023 were most pronounced in the northern hemisphere and tropics (Figs. 3b,S4a), consistent with the warming pattern (Fig. 3a). Regional maxima of the 2023 ASR anomaly, locally around 10W/m$^{-2}$, occurred over the eastern Indian Ocean, over South America and extending over the eastern Pacific in the northern branch of the inter-tropical convergence zone, as well as over northern North America, in the Southern Ocean around 60S, in the subtropical and eastern North Atlantic, and in parts of the North Pacific (Fig. 4d,S6h). All of these are present in both datasets, but the anomalies in the North Atlantic and North Pacific are more pronounced in CERES, which may be related to the handling of aerosols in ERA5 (see below).

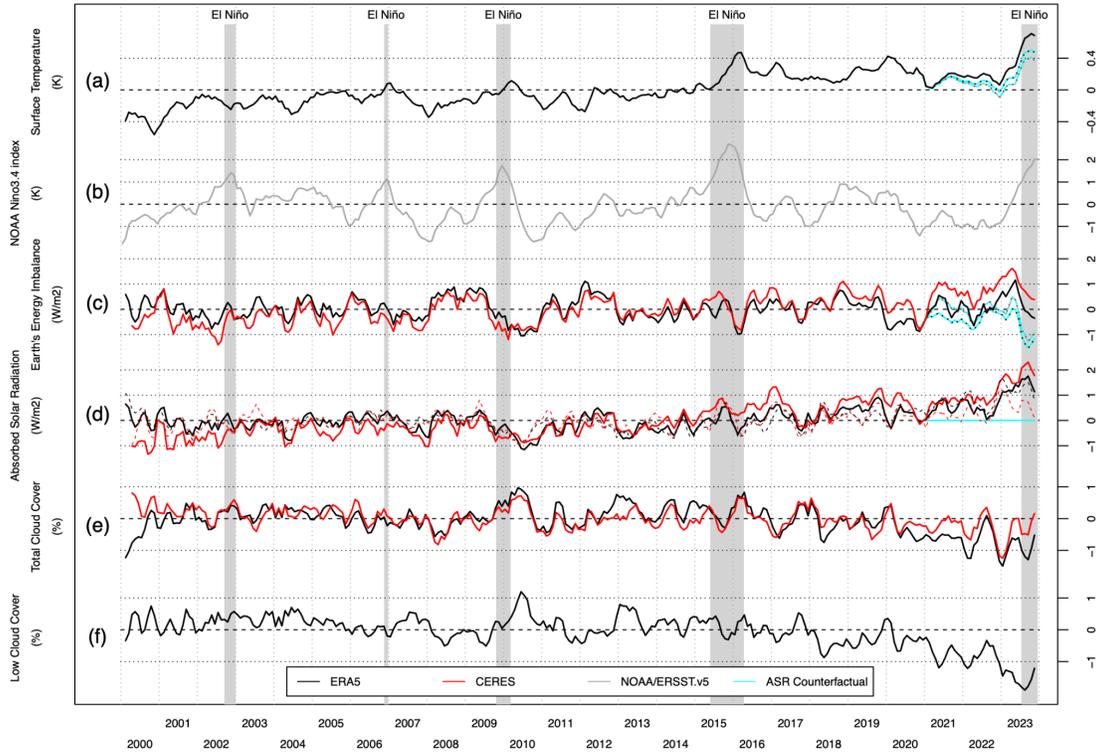

***Fig. 2. Global-mean anomalies of key parameters related to Earth's temperature, energy budget and clouds.*** *Three-month running-mean anomalies relative to 2001–2022 of (a) surface (skin) temperature, (b) NOAA Ocean Niño 3.4 index, (c) Earth's TOA total energy imbalance, (d) TOA net solar radiation (= absorbed solar radiation, ASR), (e) total cloud cover fraction, and (f) low-cloud cover fraction. Red curves show satellite data from CERES and black curves reanalysis data from ERA5. Dashed curves in (d) show the TOA solar cloud radiative effect inferred from total cloud cover anomalies (CREtc). Cyan curves show the full counterfactuals based on a 2-layer energy budget model where ASR anomalies are assumed to be zero from the beginning of December 2020 onward. El Niño periods with anomalies exceeding +1K are highlighted with gray shading. Annual means for ERA5 data starting 1940 and additional quantities are shown in Figs. S1 and S2.*

The positive ASR anomalies over the northern extratropical oceans in 2023 are broadly consistent with decadal trends prior to 2023 (Fig. 4c,d). This is not the case over the tropical Pacific and Indian Ocean, where inter-annual ASR anomalies are dominated by total cloud cover (TCC) changes associated with the El Niño–Southern Oscillation (ENSO; *Kato2009, Radley2014,Stephens2015,Timmermann2018,Loeb2024*) which transitioned from persistent La Niña to El Niño conditions after 2022 (Fig. 2b). A composite of nine El Niño events based on ERA5 data (*Methods;* Niño 3.4 index based on *Huang2017*) suggests that the albedo signature of El Niño onset years may contribute roughly 5% (+0.08W/m$^2$; Fig. S7c) to the total 2023 ASR anomaly. Regionally, ENSO-related cloud and associated ASR patterns (Fig. S7i,j) largely explain inconsistencies between 2023 anomalies and 2013–2022 trends in the tropics (Fig. 4c,d). This includes the strongly positive ASR anomalies (negative cloud anomalies) in the eastern Indian Ocean in 2023.

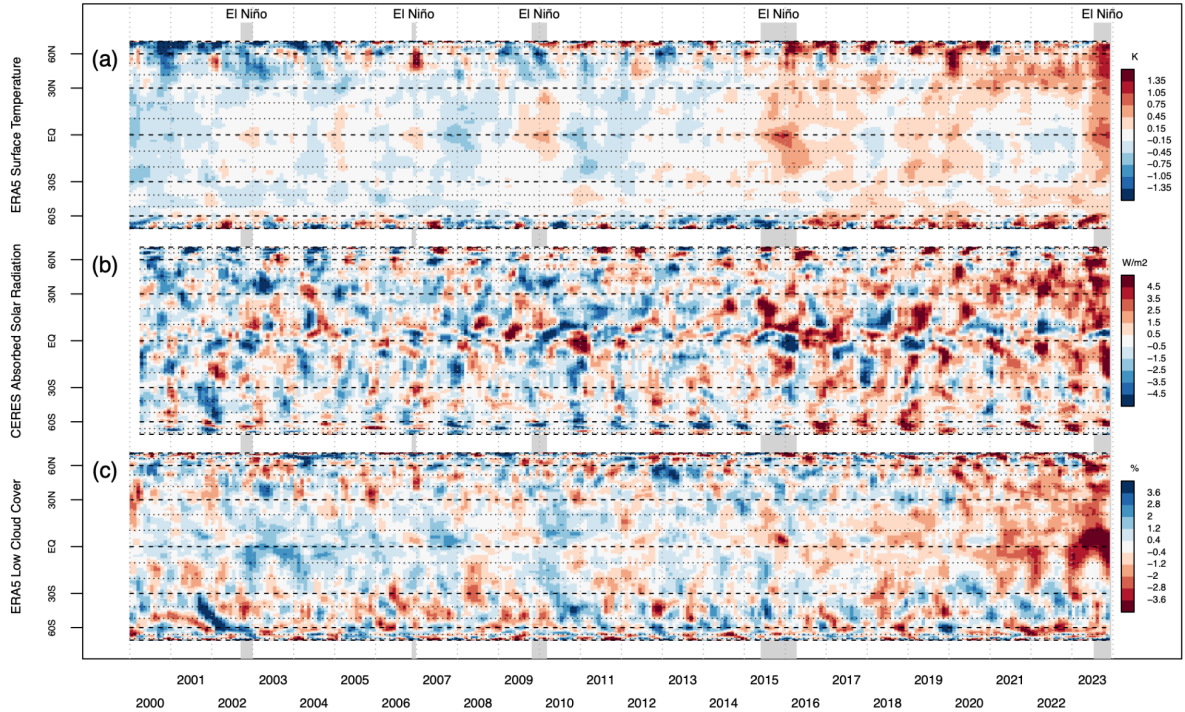

*Fig. 3. Zonal-mean anomalies of key parameters related to Earth's temperature, energy budget and clouds.* Three-monthly running-mean anomalies relative to 2001–2022 of (a) ERA5 surface (skin) temperature, (b) CERES absorbed solar radiation, and (c) ERA5 low cloud cover. El Niño periods exceeding +1K are highlighted with gray shading. Latitude spacing corresponds to cosine(latitude) for an equal-area representation. Additional parameters are shown in Fig. S4.

Before exploring cloud changes more generally, we consider the influence of surface albedo which has been declining since the 1970's (Fig. S3j), first primarily due to Arctic sea-ice and snow retreat (*Letterly2018*) and since 2016 due to Antarctic sea-ice retreat (*Riihelä2021, Roach2024*). This led to a pronounced seasonal signature in global-mean surface albedo anomalies (Fig. S2c) and polar ASR anomalies (Fig. 3b). In austral summer 2022/23, the surface albedo anomaly of -0.4% was about as strong as the planetary albedo anomaly (Fig. S2b). However, surface albedo anomalies are attenuated by about a factor 3 on average, primarily due to cloud masking (*Donohoe2011, Loeb2019*), and even more in the cloudy polar regions (*Kato2006*). Surface albedo thus contributed only weakly to the recent planetary albedo decline, in particular when averaged annually and globally, further quantified below.

**Absorbed solar radiation anomalies closely linked with cloud changes**

Given the central role of clouds in Earth's radiation budget (*Kato2009, Donohoe2011, Zelinka2012, Klein2017, Loeb2019, Loeb2021, Loeb2024*), spatial patterns of positive ASR anomalies and trends (Fig. 4c,d) and negative cloud anomalies and trends (Fig. S5a,b) are highly correlated. However, the actual influence of total cloud cover (TCC) on the shortwave cloud radiative effect (CRE) and thus ASR depends strongly on surface albedo and TOA incident solar radiation. To quantify the contribution of TCC changes to ASR anomalies, we have derived empirical relations between the two parameters for CERES and ERA5 based on 2001–2014 data, when ASR was still relatively stationary (*Methods*).

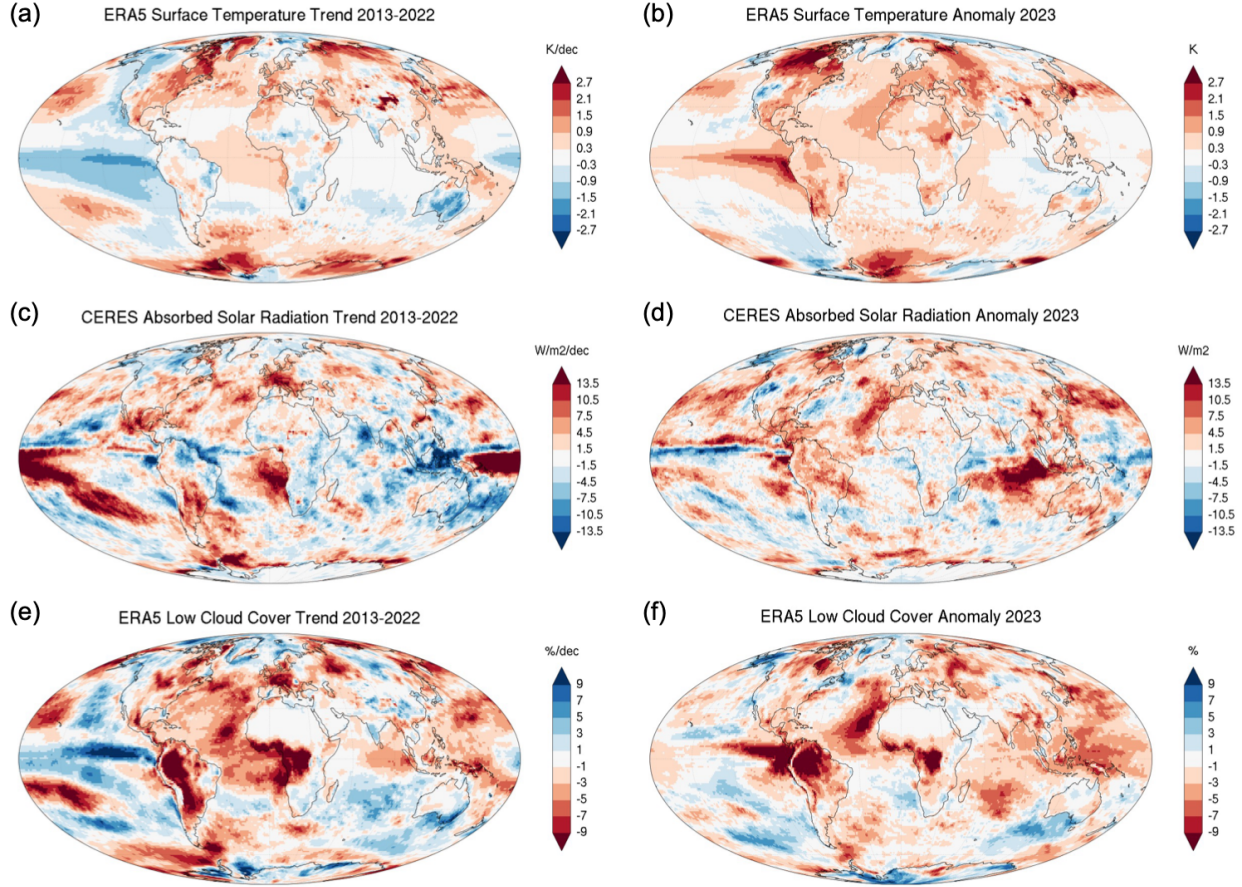

*Fig. 4. Decadal 2013–2022 trends and annual-mean 2023 anomalies of key parameters related to Earth's temperature, energy budget and clouds. Trends and anomalies relative to 2001–2022 of (a,b) ERA5 surface (skin) temperature, (c,d) CERES absorbed solar radiation and (e,f) ERA5 low-cloud cover fraction. Additional parameters are shown in Figs. S5 and S6.*

The shortwave cloud radiative effect inferred from total cloud cover (CREtc) closely resembles spatial ASR patterns (Fig. S5c,d vs. 4c,d), with the exception of the polar regions where surface albedo anomalies linked to sea-ice and snow can dominate ASR anomalies. In ERA5 (Fig. S6c,d vs. S6g,h), not just the patterns but also the magnitudes match closely, with the 2023 global-mean CREtc of +1.21W/m² close to the ASR anomaly of +1.33W/m² (Tab. 1) and with coherent temporal variations (Fig. 2d). In CERES, however, the CREtc anomalies and trends account for only about one third of the global-mean ASR anomaly and trend (Tab. 1). This is consistent with the combination of a stronger ASR trend in CERES and a weaker total cloud cover trend in CERES compared to ERA5, in particular since 2020 (Fig. 2d,e).

The different degree to which CREtc anomalies contribute to ASR anomalies can not be explained by different surface albedo contributions, which are similar between the datasets, largely constrained to high latitudes, and overall too small. Rather, it suggests that, according to CERES, either cloud reflectivity has increased beyond pre-2015 TCC-CRE relations, or clear-sky absorbed solar radiation has increased beyond the influence of surface albedo, or both. Before addressing the possible role of aerosols, which can influence both cloud and clear-sky reflectivity as well as cloud amount (*Twomey1977,Quaas2010,Li2022,Hodnebrog2024*), we consider the height-dependence of cloud trends.

**Cloud anomalies mainly due to reduced low-level clouds**

According to ERA5, the total cloud cover trends and anomalies are related mainly to declining low-level cloud cover (LCC; Figs. 2f,3c,4e,f,S3g), whereas high- and mid-level clouds have declined only slightly, if at all (Fig. S2d,e), consistent with *Loeb2024*. Regions with coherent low-cloud reductions both over 2013–2022 and in 2023 include the warm pool region around the Maritime Continent and the northern extratropical western Pacific, as well as large parts of the Atlantic and adjacent land regions. Most of these regions also exhibited reduced total cloud cover (Figs. S5a,b,S6a,b) and thus increased ASR (Figs. 4c,d,S6g,h), with the exception of the eastern part of the warm pool where changes in higher-level clouds have overcompensated the low-cloud reductions.

The South Atlantic between 20°S and the equator did not exhibit negative low-cloud anomalies in 2023, but a pronounced trend prior to 2023. Also in large parts of the East Pacific and South Indian Ocean the 2023 low-cloud anomalies were not consistent with the 2013–2022 trend. This suggests that inter-annual variability has dominated the anomalies and trends there, even though corresponding patterns in the composite of El Niño onset years (Fig. S6e) are weaker.

Globally averaged, the negative low-cloud anomaly in 2023 was about -1.5%, following a decadal trend of -1.27%/decade (Tab. 1). Relative to the absolute global-mean LCC around 38% (Fig. S1j), the 2023 anomaly amounted to astonishing -4.0%. It is striking that the eastern North Atlantic, one of the main drivers of the global-mean temperature surge (Fig. 4b; *Kuhlbrodt2024*), experienced pronounced low-cloud reductions not only in 2023 (Fig. 4f); almost the entire Atlantic experienced a substantial decline over the previous decade (Fig. 4e).

Further characterizing the cloud anomalies in and beyond 2023 based on more detailed cloud data, such as MODIS-derived cloud properties (*Minnis2021,Loeb2024*), will be important to assess the cloud trends and to further resolve differences between CERES and ERA5, including the inconsistent total cloud trends and their contribution to ASR trends. More detailed analyses will also be required to understand the causes of the observed cloud and ASR anomalies and trends, including the role of aerosols, addressed briefly in the following.

**Role of aerosols remains unclear**

There is evidence for significant aerosol contributions to recent warming trends (*Hodnebrog2024*), but isolating the contribution of indirect aerosol effects to cloud amount and reflectivity changes remains challenging. In contrast, the clear-sky absorbed solar radiation (ASR) can provide evidence for direct aerosol effects, despite confounding influences from surface albedo and atmospheric water vapor (*Taylor2007*). The sea-ice and snow retreat has led to increased global-mean clear-sky ASR (Figs. S2a,S3h) and dominates clear-sky ASR anomalies in high latitudes (Figs. S5e,f,S6e,f). Over the open ocean, however, clear-sky ASR anomalies and trends in CERES (Fig. S5e,f) may hint at direct aerosol effects, albeit these are an order of magnitude smaller compared to the corresponding all-sky values (Fig. 4c,d).

In 2023, CERES clear-sky ASR anomalies were broadly positive between the equator and about 45°N over the Atlantic and Pacific Ocean, with peaks off the East Asian Pacific coast and off the African Atlantic coast (Fig. S5f). The latter may be related to reduced transport of Saharan dust due to weakened trade winds during northern spring and summer 2023 (*Kuhlbrodt2024*), whereas the former and the broader positive signal may hint at reduced aerosols of different origin, possibly due to reduced sulfur emissions from shipping (*Diamond2023*, *Hansen2024*). Given that changed aerosol concentrations are a prerequisite for indirect aerosol effects, this also suggests that aerosols may have contributed to the reduced cloud cover and/or cloud reflectivity in these regions (*Diamond2023*, *Hansen2024*). However, while the 2013–2022 clear-sky ASR trends suggest a consistent increase along the East Asian Pacific coast, trends prior to 2023 were rather negative over the Atlantic and most of the southern-hemisphere oceans (Fig. S5e). The weak negative incident solar radiation (ISR) trend over this period (Tab. 1) is insufficient to explain these negative clear-sky ASR trends. The contribution of potentially reduced aerosol effects associated with the IMO regulations in 2015 and 2020 thus remains unclear.

Apart from the regions with sea-ice and snow retreat, clear-sky ASR anomalies are much weaker and smoother in ERA5 (Fig. S6e,f) where aerosols are prescribed (*Hersbach2020*) based on forcing data of the Coupled Model Intercomparison Project (CMIP). Alongside the recent ISR underestimation, this may add to the lower 2023 ASR and imbalance anomalies in ERA5 relative to CERES (Tab. 1), in particular over the North Pacific and North Atlantic.

**Temperature response to reduced planetary albedo can explain the warming gap**

Earth's climate responds to different types of forcing in a complex way (*Hansen2005*), but it is possible to estimate the influence of the record-low planetary albedo and the associated absorbed shortwave radiation (ASR) anomalies on global-mean surface temperature (GMST) with a two-layer energy budget model (EBM; *Held2010, Geoffroy2013*). To construct counterfactuals where ASR anomalies are assumed to be zero from some time onward, we integrate monthly CERES and ERA5 ASR anomalies relative to 2001–2022 until December 2023, starting in December each year from 2001 to 2022 (*Methods*; Figs. S10,S11). Subtracting the EBM upper-layer temperature response from the observed GMST provides counterfactual realizations of how GMST may have evolved without the ASR anomalies; counterfactuals of the imbalance (EEI) are constructed in the same way. Additional counterfactuals are constructed with ASR anomalies only from different zonal bands and from incident solar radiation. Considering only the shortwave perturbations is reasonable given that the planetary albedo decline is associated primarily with low-level clouds, which lack the compensating longwave effects of mid- and high-level clouds (*Zelinka2016*).

We focus on the counterfactuals starting December 2020, about the last time when ASR anomalies in satellite data were close to zero and when the clearest low-cloud and ASR trends set in (Fig. 2d,f). Based on the full counterfactuals, the 2023 annual-mean GMST may have been 0.25K cooler based on the CERES counterfactual and 0.16K cooler based on the ERA5 counterfactual (cyan curves in Fig. 2a). Subtracting 0.026K due to the solar intensity increase after December 2020 (Fig. S2f) in CERES, not captured in ERA5 (-0.016K), the discrepancy is reduced. The full counterfactual total imbalance (EEI) anomaly drops below -1W/m$^2$ towards the end of 2023 (Fig. 2c). Again adjusting for the solar intensity results in less negative counterfactual EEI anomalies in both cases, more consistent with previous El Niño events.

Importantly, the CERES estimate for the effect of plenary albedo alone on the 2023 GMST is about 0.23K (Fig. 1b), approximately the magnitude of unexplained warming in 2023.

The polar regions beyond 55S and 55N, where surface albedo decline due to sea-ice and snow retreat strongly affects the ASR, jointly contribute about 0.03K (12%) to the ASR-driven warming response (Fig. 1b). The remaining 0.20K are dominated by the northern mid-latitudes and tropics where cloud changes drive ASR anomalies. By comparison, based on the Niño 3.4 index in 2023 compared to previous El Niño onset years and their annual-mean GMST anomalies (*Methods*), we estimate that El Niño has contributed only about +0.07K to the 2023 temperature anomaly, leaving a residual of +1.15K warming that may have occurred in 2023 without the anomalous solar absorption and El Niño (Figs. 1b,S8). The total GMST response of +0.23K to the reduced planetary albedo found here is broadly consistent with the amount of cooling observed after major volcanic eruptions relative to the respective forcing (Fig. S3a,d; *Hansen2005,Zanchettin2022*), adding credibility to our results.

**Major potential implications for climate sensitivity**

Three fundamental mechanisms may have contributed to the record-low planetary albedo associated with reduced low-level clouds: internal variability, an emerging low-cloud feedback, and aerosol effects. Contributions from internal variability would subside and leave our expectation of the longer-term warming unaffected. The relative stationarity of the low-cloud cover until about 2015 (Figs. 2f and 3c) speaks against short-term variability, but longer-term variability associated for example with the Atlantic Multidecadal Variability (AMV; *Kuhlbrodt2024,Li2024*) could contribute to the observed trends, also given that ocean surface warming can reduce low-cloud cover (*Klein2017,Boehm2023,Athanase2024*).

The latter mechanism is also essential if the recent trends are due to an emerging low-cloud feedback unrelated to internal variability, complicating a separation of the two. The response of low clouds is the largest source of uncertainty driving differences in climate sensitivity between climate models (*Qu2014*), even after the expected range of low-cloud response and climate sensitivity could be reduced with observational constraints (*Klein2017,Sherwood2020*). If a substantial low-cloud feedback now emerges in observations, the lower end of realistic climate sensitivity estimates may need to be adjusted upward.

The average combined shortwave forcing by aerosols in CMIP6 climate models was -1.26W/m$^2$ (*Smith2020*). A near-complete loss of anthropogenic aerosols would thus be required to match the observed ASR anomaly in 2023, speaking either for an underestimated aerosol effect in models (*Hansen2024*) or strong contributions from internal variability or low-cloud feedback (*Loeb2024*). Even though the negative correlation between the aerosol effect and climate sensitivity found in CMIP3 (*Kiehl2007*) was weaker in later CMIPs (*Forster2013,Smith2020*), a stronger historical aerosol cooling would require a higher sensitivity to greenhouse forcing to reproduce the observed temperature record.

In summary, if the cloud-related albedo decline was caused not solely by internal variability, the 2023 extra heat may be here to stay and Earth's climate sensitivity may be closer to the upper range of current estimates. We may thus be closer to the temperature targets defined in the Paris agreement than previously thought, with strong implications for remaining carbon budgets.

**Acknowledgments:** We thank Hans Hersbach and Adrian Simmons for answering our questions about the interpretation of ERA5 trends and aerosol handling. Sebastian Bathiany, Tobias Becker, Suvarchal Cheedala and Antje Boetius provided very valuable comments on a draft version of this paper. We are grateful to the Copernicus Climate Change Service (C3S) and the National Aeronautics and Space Administration (NASA) for the development, production and provision of the invaluable CERES-EBAF and ERA5 data sets. We thank the German Climate Computing Centre (DKRZ) for providing simple access to the ERA5 data and computing resources for the data analysis.

**Funding:** HFG and TJ based on internal funds of the Alfred Wegener Institute, Helmholtz Centre for Polar and Marine Research, which is part of the Helmholtz Association and funded by the Germany Ministry of Education and Research (BMBF). This research has been supported by the European Commission Horizon 2020 Framework Programme nextGEMS (grant no. 101003470).

**Author contributions:** Conceptualization: HFG; Methodology: HFG, TR; Investigation: HFG, TR, TJ; Visualization: HFG, TR; Funding acquisition: /; Project administration: /; Supervision: /; Writing – original draft: HFG, TR, TJ; Writing – review & editing: HFG, TR, TJ

**Competing interests:** Authors declare that they have no competing interests.

**Data and materials availability:** ERA5 data is available from the Copernicus Climate Change Service (C3S) Climate Data Storage at https://cds.climate.copernicus.eu/. CERES-EBAF data is available from NASA at https://asdc.larc.nasa.gov/project/CERES. NOAA Climate Prediction Center Ocean Niño Index data is available at https://www.cpc.ncep.noaa.gov/data/indices/oni.ascii.txt.

**Supplementary Materials**

    Materials and Methods

        Data

        Temperature, cloud and albedo signatures of El Niño

        Empirical estimation of total cloud cover-inferred ASR anomalies

        2-layer energy balance model and ASR-based counterfactuals

    Figs. S1 to S11

    Table S1

# Supplementary Materials for

## Recent global temperature surge amplified by record-low planetary albedo

Helge F. Goessling, Thomas Rackow, and Thomas Jung

Corresponding author: helge.goessling@awi.de

**The PDF file includes:**

Materials and Methods

    Data

    Temperature, cloud and albedo signatures of El Niño

    Empirical estimation of total cloud cover-inferred ASR anomalies

    2-layer energy balance model and ASR-based counterfactuals

Figs. S1 to S11

Table S1

**Materials and Methods**

Data

We use three public datasets to explore Earth's energy budget, surface temperature and clouds, namely (i) the Clouds and the Earth's Radiant Energy System (CERES) Energy Balanced and Filled (EBAF) Top-of-Atmosphere (TOA) Edition-4.2 Data Product from the US National Aeronautics and Space Administration (NASA; *Loeb2018*, hereafter CERES), (ii) the ERA5 reanalysis produced by the European Centre for Medium-Range Weather Forecasts (ECMWF; *Hersbach2020,Bell2021*), and the Ocean Niño 3.4 Index (ONI) produced by the US National Oceanic and Atmospheric Administration (NOAA).

CERES is a satellite-based product intended to provide a long-term record for detecting decadal changes in Earth's global radiation budget, clouds and aerosols (*Loeb2018*). Besides measurements of radiation emitted from Earth, CERES also includes measurements of solar intensity. CERES includes a one-time adjustment of the TOA fluxes to ensure that the global-mean net TOA flux for July 2005–June 2015 is consistent with an in-situ value of 0.71 $Wm^{-2}$. In addition to the monthly TOA all-sky and clear-sky radiation fluxes on a 1-degree longitude-latitude grid, available from March 2000–December 2023, we use the total cloud cover fractions based on the Moderate Resolution Imaging Spectrometer (MODIS) contained in the CERES product (*Minnis2021*).

ERA5 is ECMWF's latest reanalysis product, updated on a daily basis in near-real-time (*Hersbach2020,Bell2021*). It combines an extensive set of satellite and in-situ observational data with a fixed version of ECMWF's physical numerical weather prediction model by a sophisticated data assimilation scheme. It thereby provides a gap-free global gridded estimate of the evolution of a large number of variables, including clouds and radiation fluxes, reaching back to January 1940. Long-term trends in ERA5 need to be interpreted with caution due to possible shifts resulting from changes in the observing system, including strongly reduced data density in earlier decades. Given that CERES data is also subject to uncertainties, we use ERA5 (i) as an additional data product, even though not fully independent as the satellite data sources are overlapping, (ii) to get some evidence for the evolution of relevant parameters before March 2000, keeping in mind the possibility of spurious trends, and (iii) to analyze the height-dependence of cloud changes based on three categories, namely low-level (atmospheric pressure above about 800hPa), mid-level (atmospheric pressure between about 800hPa and 450hPa), and high-level (atmospheric pressure below about 450hPa) clouds. For consistency, all monthly ERA5 data has been remapped from the original 0.28-degree grid to the 1-degree CERES grid by first-order conservative remapping prior to analysis.

Total cloud cover is about 4-5% lower in ERA5, although this may be due to differences in cloud definitions rather than being an actual bias. Apart from that, global-mean ERA5 and CERES climatologies are matching rather closely (Fig. S1). Generally, spatio-temporal variations are consistent between the datasets, as visible in numerous figures in this study, suggesting that ERA5 is suitable for the purposes listed above.

The NOAA Niño 3.4 index is constructed from 3-month running means of ERSST.v5 sea-surface temperature anomalies in the Niño 3.4 region (5S–5N, 170W–120W), based on centered 30-year base periods updated every 5 years (*Huang2017*).

Temperature, cloud and albedo signatures of El Niño

El Niño is a key driver of inter-annual climate variability (*Radley2014,Timmermann2018*). Given that conditions in 2023 transitioned from La Niña to El Niño (Fig. 2b), we estimate the impact of El Niño on temperatures, radiation fluxes and clouds in 2023 based on previous El Niño events. We consider a set of the 4 pre-2023 El Niño events covered by the CERES period (2002/03, 2006/07, 2009/10, 2015/16) based on CERES and ERA5 data, and an extended set of 9 pre-2023 El Niño events (1951/52, 1957/58, 1968/69, 1972/73, 1997/98, 2002/03, 2006/07, 2009/10, 2015/16) based on ERA5 data. Events influenced by volcanic eruptions and the double-event 1987/88 (Fig. S3) are excluded. To isolate anomalies associated only with El Niño, anomalies for each event have been normalized by subtracting the average anomalies of the two years flanking each event, that is, the year before the El Niño onset year and the year after the El Niño second year. Nonlinear longer-term trends, including seasonally-varying ones, are thereby removed. We consider multi-event composites to estimate the temporal evolution of several parameters over the course of an average event (Fig. S7), including the shortwave signature, and individual events to estimate the temperature anomaly due to El-Niño in 2023 (Fig. S8). Given that 2023 was an El Niño onset year, the main focus is on these.

During El Niño, warmer ocean surface waters are exposed in the eastern tropical Pacific by internal redistribution of water masses (*Timmermann2018*; Fig. S7g), resulting in higher global-mean surface temperature (Figs. 2a,b,S7a). During previous events, the normalized annual-mean GMST anomaly during El Niño onset years was mostly around +0.05K to +0.10K, with a dependence on the annual-mean Niño 3.4 index (Fig. S8). Based on the 2023-mean Niño 3.4 index anomaly and linear least-squares regression, we estimate that El Niño has contributed about +0.07K to the 2023 temperature anomaly.

Modulated by a complex interplay of different energy fluxes and mechanisms (*Zeppetello2019*), the surface warming associated with El Niño leads to increased global-mean TOA outgoing longwave radiation. This results in a decreased total imbalance (EEI) (Fig. S7b, consistent with *Radley2014*), although modified by a total absorbed solar radiation (ASR) signature (Fig. S7c,h) and hence a planetary albedo signature (Fig. S7d) related mainly to cloud patterns (Fig. S7e,f,i,j) (*Kato2009,Radley2014,Stephens2015*). Based on the ERA5-based nine-event composite, the global and annual-mean ASR signature during El Niño onset years is +0.08Wm$^{-2}$ (planetary albedo -0.02%; dashed curves in Fig. S7c,d) and thus only a small fraction of the total 2023 ASR anomaly of +1.82Wm$^{-2}$ in CERES and +1.33Wm$^{-2}$ in ERA5. The contribution of the El Niño albedo signature to the 2023 GMST anomaly is about an order of magnitude smaller compared to the general temperature signature of +0.07K (see above). The possible ASR contribution of El Niño is thus not treated separately in counterfactuals based on the energy balance model integrations (see below).

Empirical estimation of total cloud cover-inferred ASR anomalies

To estimate the contribution of total cloud cover (TCC) changes to the ASR anomalies and trends, we have fitted power functions of the form

(1) $CRE = \alpha \cdot TCC^\beta$

to the monthly-mean TCC and shortwave cloud radiative effect (CRE) data for each of the 1-degree grid points for each calendar month and dataset separately. To keep the parameters within physically reasonable bounds even where the TOA incident solar radiation is very small or CRE variations are strongly confounded by surface albedo variations, we bounded $\alpha$ within [-ISR,ISR] and $\beta$ within [1,4]. We have used only the inter-annual variations within 2001–2014, because during this period global-mean ASR and low-cloud cover were still relatively stationary (Fig. 2d,f) and in order to exclude any potential aerosol-cloud effects that might have been caused by the two IMO regulation changes in 2015 and 2020. Given the small sample size, for each grid point we have also included the data of neighboring grid cells within 2 degrees in meridional and zonal direction, as long as the climatological surface albedo differs by at most 5%.

Results of fitted parameters and explained variance are shown exemplarily for May in Fig. S9. The fitted power functions have subsequently been used to infer a hypothetical CRE going back to TCC-anomalies alone, termed CREtc (Figs. 2d,S1f,S4b,d,S5c,d,S6c,d, Tab. 1). CREtc accounts not only for an isolated cloud amount effect, where CRE is proportional to TCC (*Zelinka2012*), but also for changes in CRE through covariances between TCC and cloud optical depth and altitude. We find that, arguably due to these covariances, cloud radiative effect typically depends superlinearly on total cloud cover in both datasets (Fig. S9c,d), with regional differences that may be related to the dependence of cloud overlap on cloud regimes.

2-layer energy balance model and ASR-based counterfactuals

The 2-layer energy balance model (EBM) used in this study to estimate temperature responses to absorbed solar radiation (ASR) anomalies follows earlier work by Held et al. (2010) and Geoffroy et al. (2013). The EBM adopted here splits the Earth system into two main layers. It describes the evolution of upper-layer temperature perturbations $T$ and a deep-ocean layer temperature perturbation $T_0$ over time as a system of two ordinary differential equations (ODEs)

(2) $C \frac{dT(t)}{dt} = F(t) - \lambda t - \gamma(T(t) - T_0(t))$

(3) $C_0 \frac{dT_0(t)}{dt} = \gamma(T(t) - T_0(t))$,

where $C$ is a heat capacity for the atmosphere-land-upper-ocean system; $F$ is a radiative forcing amplitude function that may vary over time, $t$; $\lambda$ is the radiative feedback parameter for a $CO_2$ perturbation; $\gamma$ is an exchange coefficient between the upper and deeper ocean layers; and $C_0$ is the deep-ocean heat capacity. The parameters of the EBM follow Table 3 and 4 in Geoffroy et al. (2013), where parameter fits to CMIP5 models are presented. We use the CMIP5 multi-model mean (MMM, based on 15 models) for the individual parameters, as repeated for convenience in Table S1. Solar forcing is less effective than an equivalent $CO_2$ forcing, with an efficacy of $F_{eff} = 92\%$ (Hansen et al. 2005). Originally this estimate is for ASR anomalies due to solar intensity anomalies rather than planetary albedo anomalies, but we assume that it holds for the latter, too. We thus multiply by this factor in $F$ when using ASR anomalies from ERA5 or CERES as input to the EBM.

Another way to look at the system of equations is that the left-hand sides of the ODEs describe the tendencies of upper-layer (Eq. 2) and deep-layer (Eq. 3) heat contents (Geoffroy et al., 2013).

The chosen MMM value for the upper-layer $C$ heat content roughly implies an effective mixed layer thickness of 77m (Geoffroy et al., 2013). With the chosen values, the upper and deeper ocean layers respond on a timescale of 4.1 years and 219 years, respectively (Geoffroy et al., 2013).

The 2-layer EBM has simple analytical solutions for linear and step-wise forcing functions $F$ (Geoffroy et al., 2013). In our implementation in Python for use with general forcing functions $F$ (e.g. from ERA5 and CERES), the system of ODEs is solved numerically with a Forward Euler discretisation and monthly timesteps. For a simple step-wise forcing, our numerical solution agrees with the analytical solutions given by Geoffroy et al. (2013). The EBM returns monthly $T$, $T_0$, and Earth Energy Imbalance (EEI), $F - \lambda t$, as output.

To construct ASR-based counterfactuals where ASR anomalies are assumed to be zero from some time onward, we drive the EBM with monthly CERES and ERA5 ASR anomalies relative to 2001–2022 until December 2023, starting in December each year from 2001 to 2022 (Figs. S10,S11). Subtracting the EBM upper-layer temperature response from the observed GMST provides counterfactuals of how GMST may have evolved without the ASR anomalies. Counterfactuals of the imbalance (EEI) are constructed in the same way. Considering only the shortwave perturbations is reasonable given that the planetary albedo decline is associated primarily with low-level clouds, which lack the compensating longwave effects of mid- and high-level clouds (*Zelinka2016*).

Additional counterfactuals are constructed by separately using the contributions to the global-mean ASR anomalies (i) from five different zonal bands (Antarctic 90S–55S; Southern mid-latitudes 55S–23S; Tropics 23S–23N; Northern mid-latitudes 23N–55N; Arctic 55N–90N) and (ii) from anomalies of the TOA incident solar radiation (Fig. S2f) by multiplication with the spatio-temporal pattern of absolute planetary albedo. Case (i) also serves to estimate the contribution of surface albedo trends to the global planetary albedo trend. This is reasonable because, first, surface albedo trends are dominated by sea-ice and snow retreat poleward of 55S and 55N and, second, surface albedo trends dominate total planetary albedo trends there.

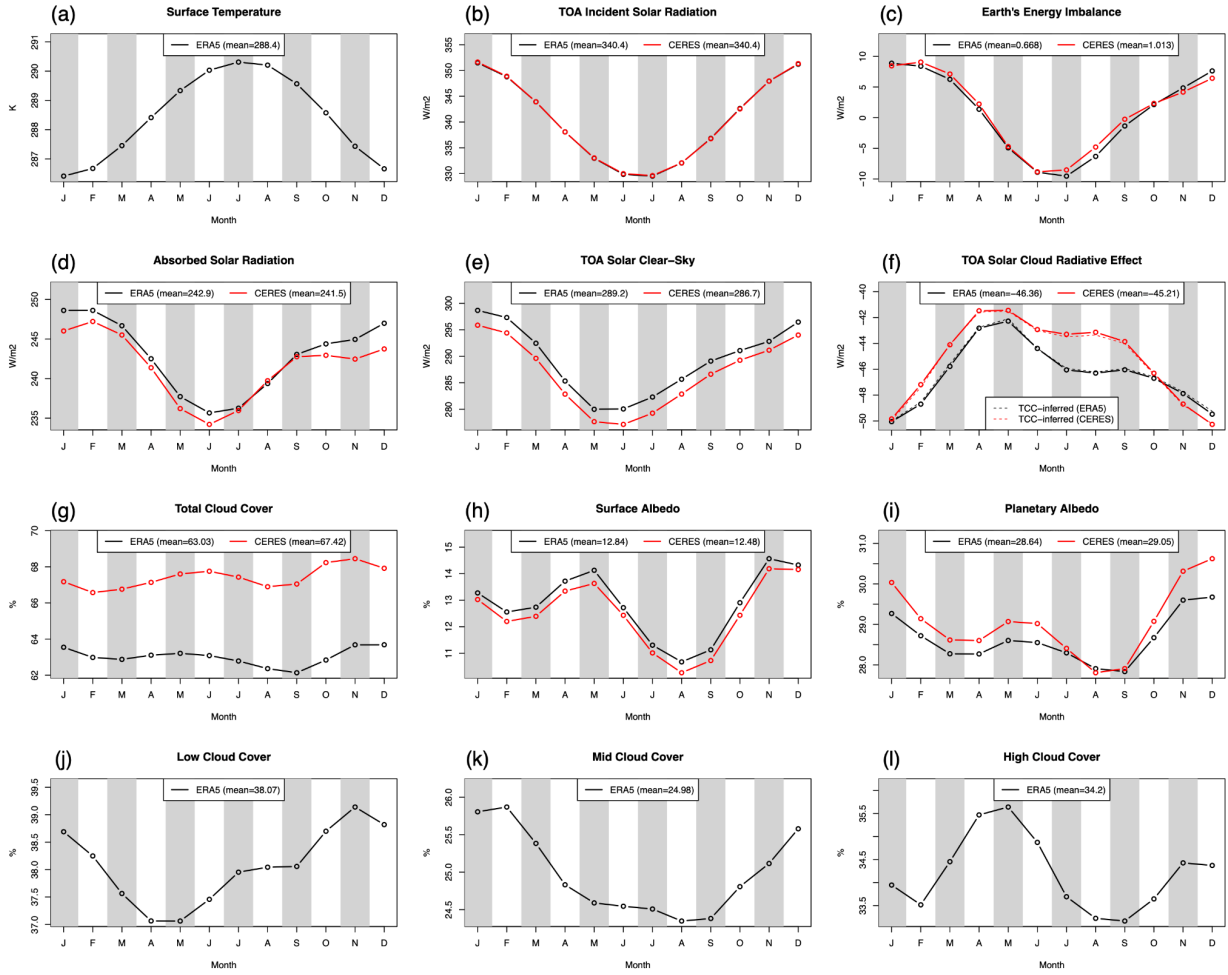

*Fig. S1. Monthly climatologies of relevant quantities related to Earth's energy budget for 2001–2022.* *(a) Surface (skin) temperature, (b) TOA incident solar radiation (ISR), (c) Earth's TOA total energy imbalance (EEI), (d) TOA net solar radiation (= absorbed solar radiation, ASR), (e) TOA net solar clear-sky radiation, (f) TOA net solar cloud radiative effect (CRE, solid) with values inferred from total cloud cover dashed (CREtc), (g) total cloud cover fraction, (h) surface albedo (derived from global-mean surface solar downwelling and upwelling radiation), (i) planetary albedo (derived from global-mean TOA incident solar and upwelling solar radiation), (j) mid-level cloud cover fraction, (k) mid-level cloud cover fraction, and (l) high-level cloud cover fraction. Red curves show CERES data and black curves show ERA5 data.*

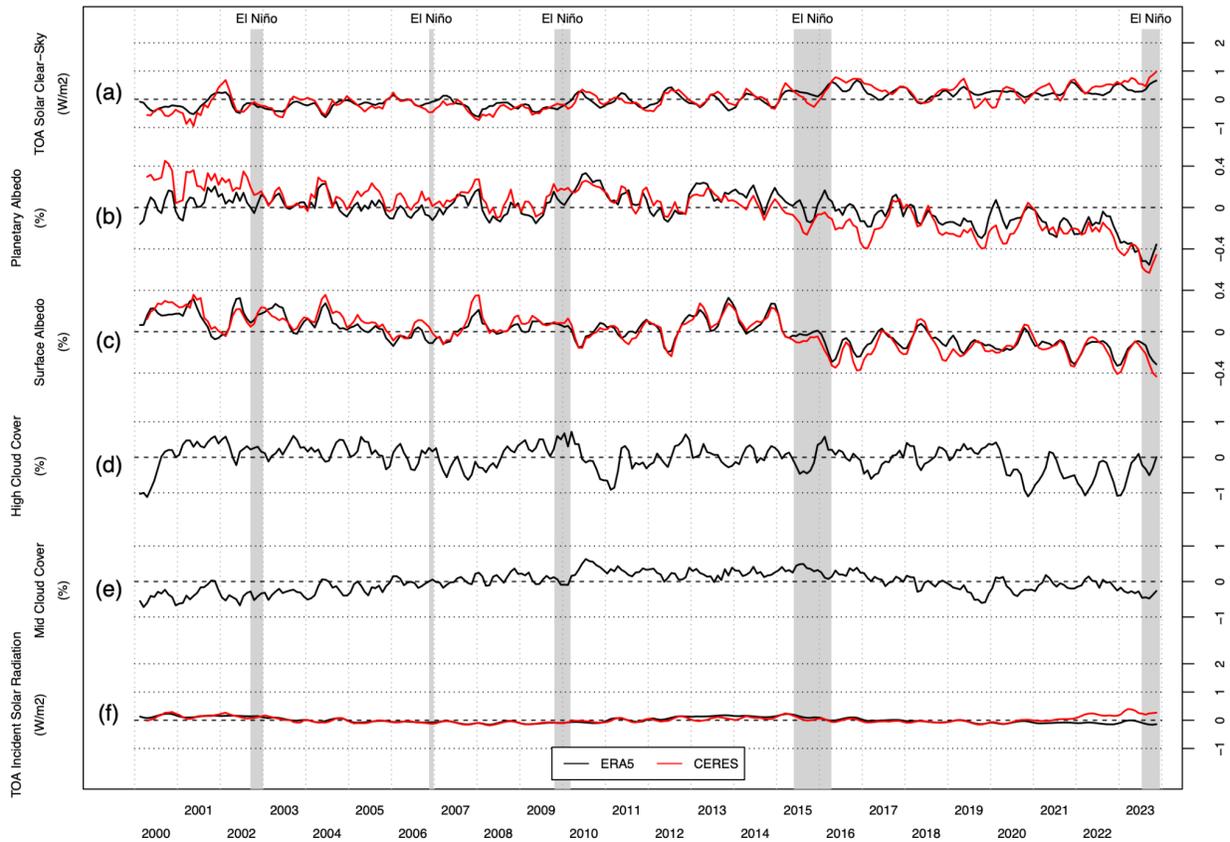

*Fig. S2. Global-mean anomalies of additional quantities related to Earth's energy budget.* Three-monthly running-mean anomalies relative to 2001–2022 of (a) TOA net solar clear-sky radiation, (b) planetary albedo (derived from global-mean TOA incident solar and upwelling solar radiation), (c) surface albedo (derived from global-mean surface solar downwelling and upwelling radiation), (d) high-level cloud cover fraction, (e) mid-level cloud cover fraction, and (f) TOA incident solar radiation. Red curves show CERES data and black curves show ERA5 data. El Niño periods with anomalies exceeding +1K (see Fig.1b) are highlighted with gray shading.

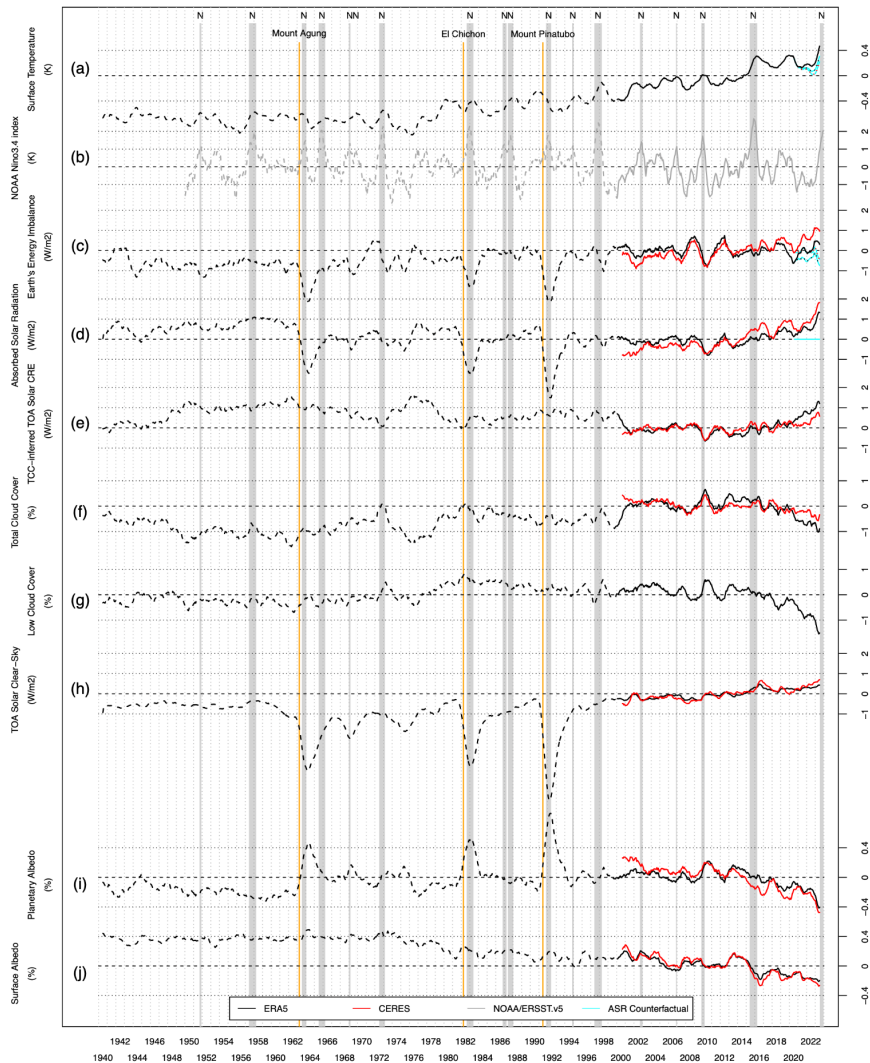

*Fig. S3. Global-mean anomalies of selected quantities related to Earth's energy budget since 1940.* Twelve-monthly running-mean anomalies relative to 2001–2022 of (a) surface (skin) temperature, (b) NOAA Ocean Niño 3.4 index, (c) Earth's TOA total energy imbalance, (d) TOA net solar radiation (= absorbed solar radiation, ASR), (e) TOA solar cloud radiative effect inferred from total cloud cover anomalies, (f) total cloud cover fraction, (g) low-level cloud cover fraction, (h) TOA net solar clear-sky radiation, (i) planetary albedo (derived from global-mean TOA solar downwelling and upwelling radiation), and (j) surface albedo (derived from global-mean surface solar downwelling and upwelling radiation). Red curves show CERES data and black curves show ERA5 data. Cyan curves show counterfactuals based on a 2-layer energy budget model where ASR anomalies are assumed to be zero from the beginning of December 2020 onward. El Niño periods with anomalies exceeding +1K are highlighted with gray shading and labeled "N". Curves showing ERA5 data before 2000 are dashed to express that long-term trends can be spurious due to observing-system changes.

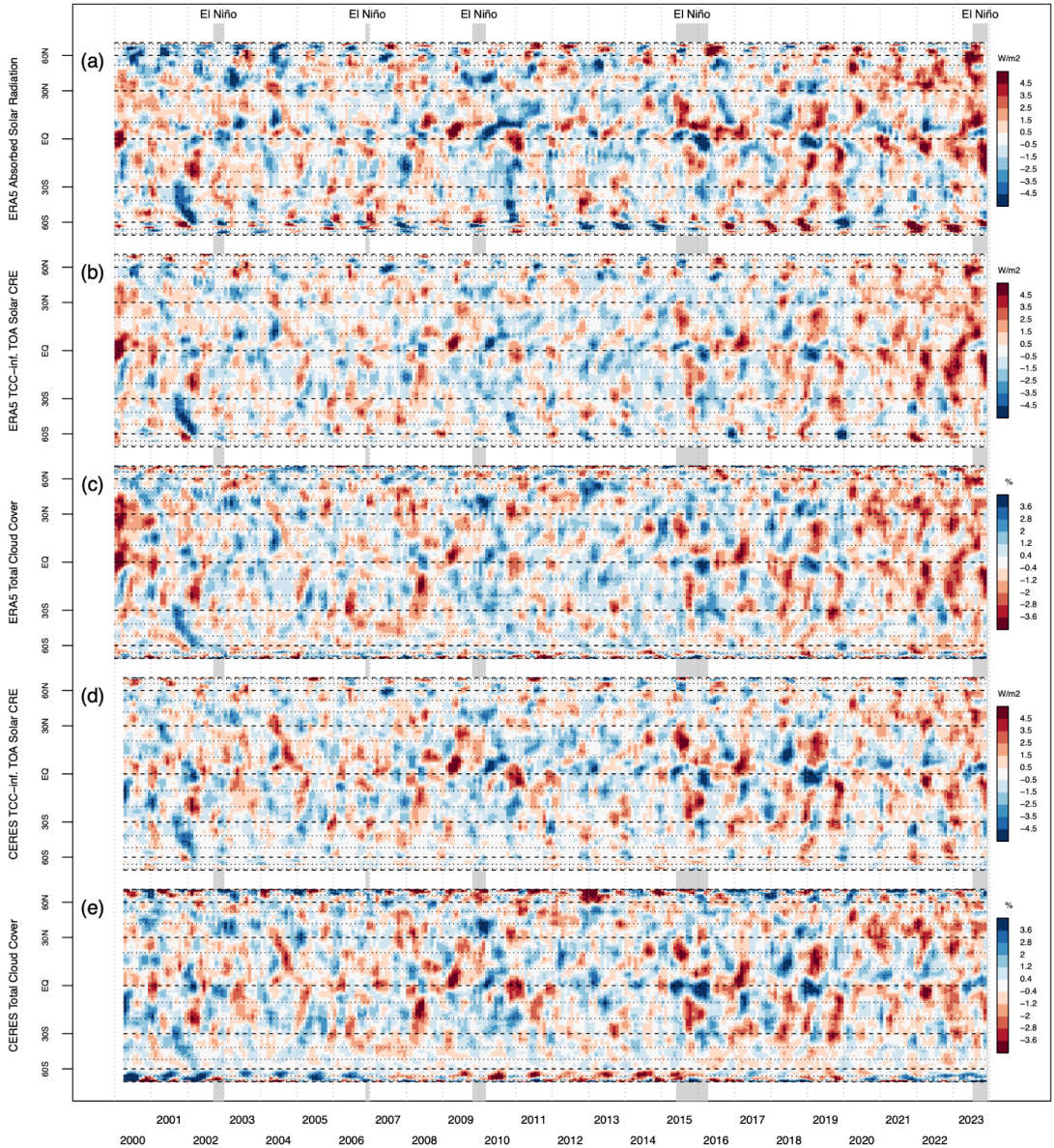

**Fig. S4. Zonal-mean anomalies of additional quantities related to Earth's energy budget.** *Three-monthly running-mean anomalies relative to 2001–2022 of (a) ERA5 absorbed solar radiation, (b) ERA5 cloud radiative effect inferred from total cloud cover anomalies, (c) ERA5 total cloud cover, (d) CERES TOA solar cloud radiative effect inferred from total cloud cover anomalies, and (e) CERES total cloud cover. El Niño periods with anomalies exceeding +1K are highlighted with gray shading. Latitude spacing corresponds to cosine(latitude) for an equal-area representation.*

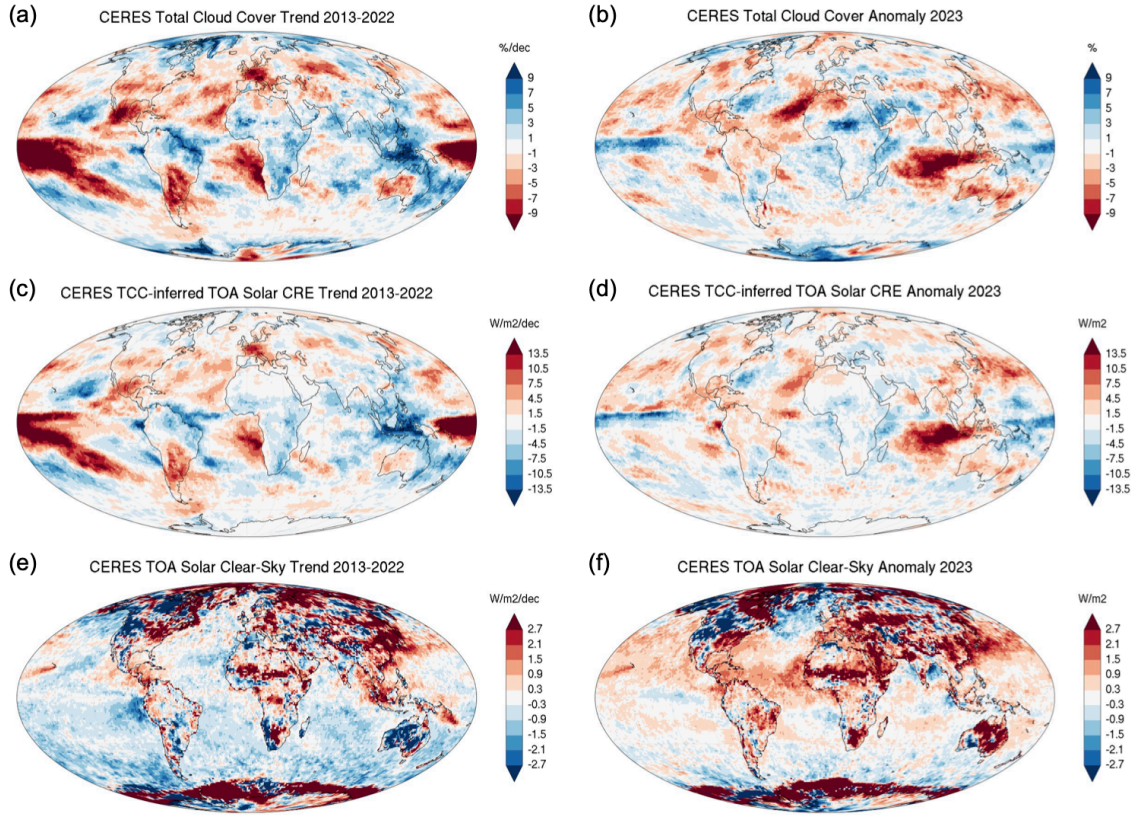

*Fig. S5. Decadal 2013–2022 trends and annual-mean 2023 anomalies of additional parameters related to Earth's energy budget and clouds.* CERES trends and anomalies relative to 2001—2022 of (a,b) total cloud cover fraction, (c,d) TOA solar cloud radiative effect inferred from total cloud cover anomalies and (e,f) TOA net solar clear-sky radiation (color scale chosen to resolve anomalies over the open ocean).

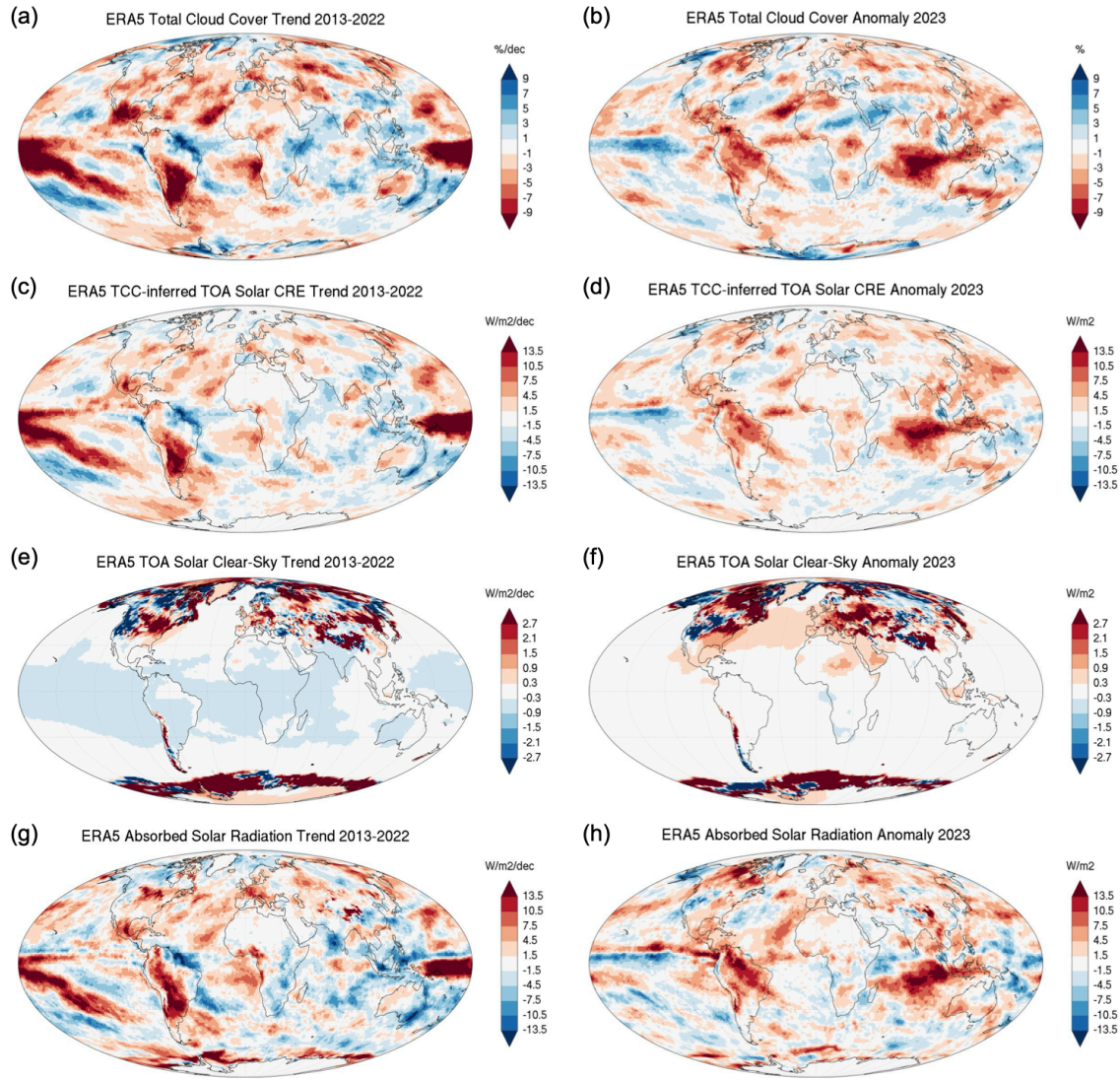

*Fig. S6. Decadal 2013–2022 trends and annual-mean 2023 anomalies of additional parameters related to Earth's energy budget and clouds. ERA5 trends and anomalies relative to 2001—2022 of (a,b) total cloud cover fraction, (c,d) TOA solar cloud radiative effect inferred from total cloud cover anomalies, (e,f) TOA net solar clear-sky radiation (color scale chosen to resolve anomalies over the open ocean) and (g,h) TOA net solar radiation (= absorbed solar radiation, ASR).*

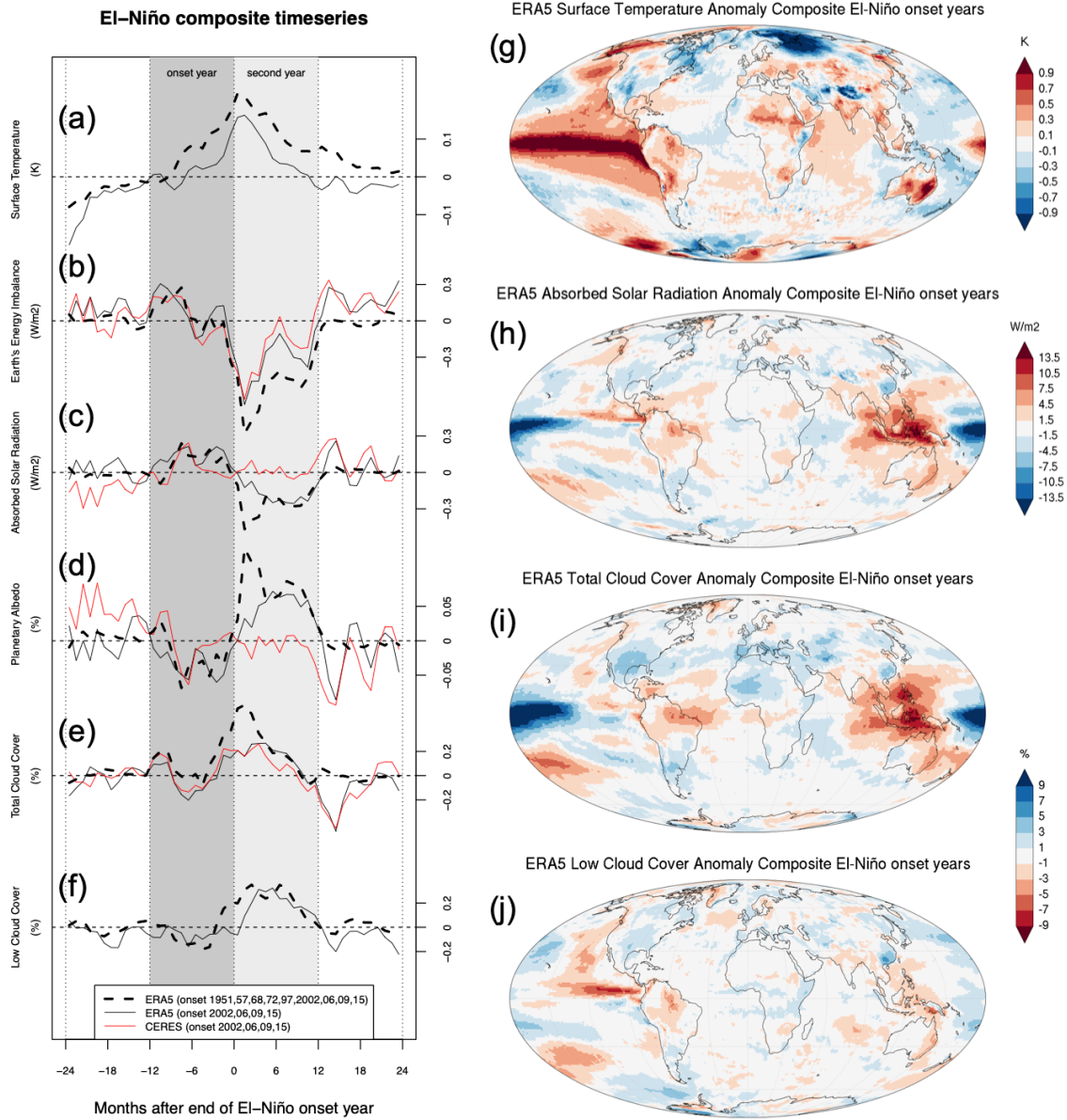

*Fig. S7. El Niño composite analysis. (a-f) Global-mean composites of 4 pre-2023 El Niño events covered by the CERES period based on CERES data (thin solid red curves) and ERA5 data (thin solid black curves) and of an extended set of 9 pre-2023 El Niño events based on ERA5 data (thick dashed black curves) for (a) surface (skin) temperature, (b) Earth's TOA total energy imbalance, (c) TOA net solar radiation (= absorbed solar radiation, ASR), (d) planetary albedo (derived from global-mean TOA solar downwelling and upwelling radiation), (e) total cloud cover fraction, and (f) low-level cloud cover fraction. (g-j) Annual-mean composites of 9 pre-2023 El Niño onset years based on ERA5 data (corresponding to the thick dashed black curves in a-f) for (g) surface (skin) temperature, (h) TOA net solar radiation (= absorbed solar radiation, ASR), (i) total cloud cover fraction, and (j) low-level cloud cover fraction.*

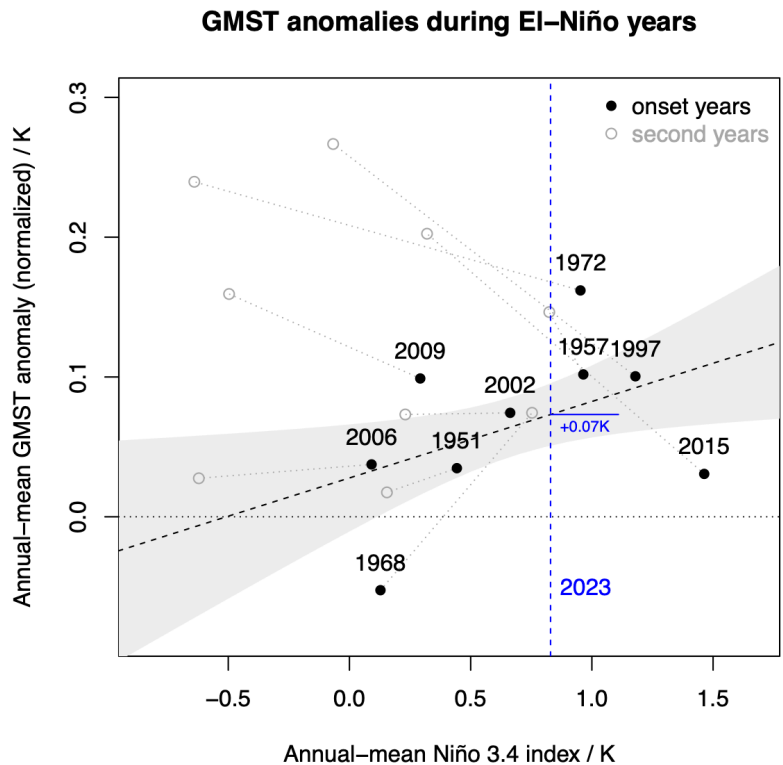

*Fig. S8. **Global-mean surface temperature anomalies during El Niño years.** Based on annual-means of NOAA Ocean Niño 3.4 index and ERA5 annual-mean surface (skin) temperature for 9 El Niño onset years (black filled circles) and corresponding second years (grey circles). The contribution of El Niño to the 2023 temperature anomaly (+0.07K) is estimated based on the 2023-mean Niño 3.4 index (blue dashed line) with linear least-square regression (black dashed line and 68% confidence band by gray shading).*

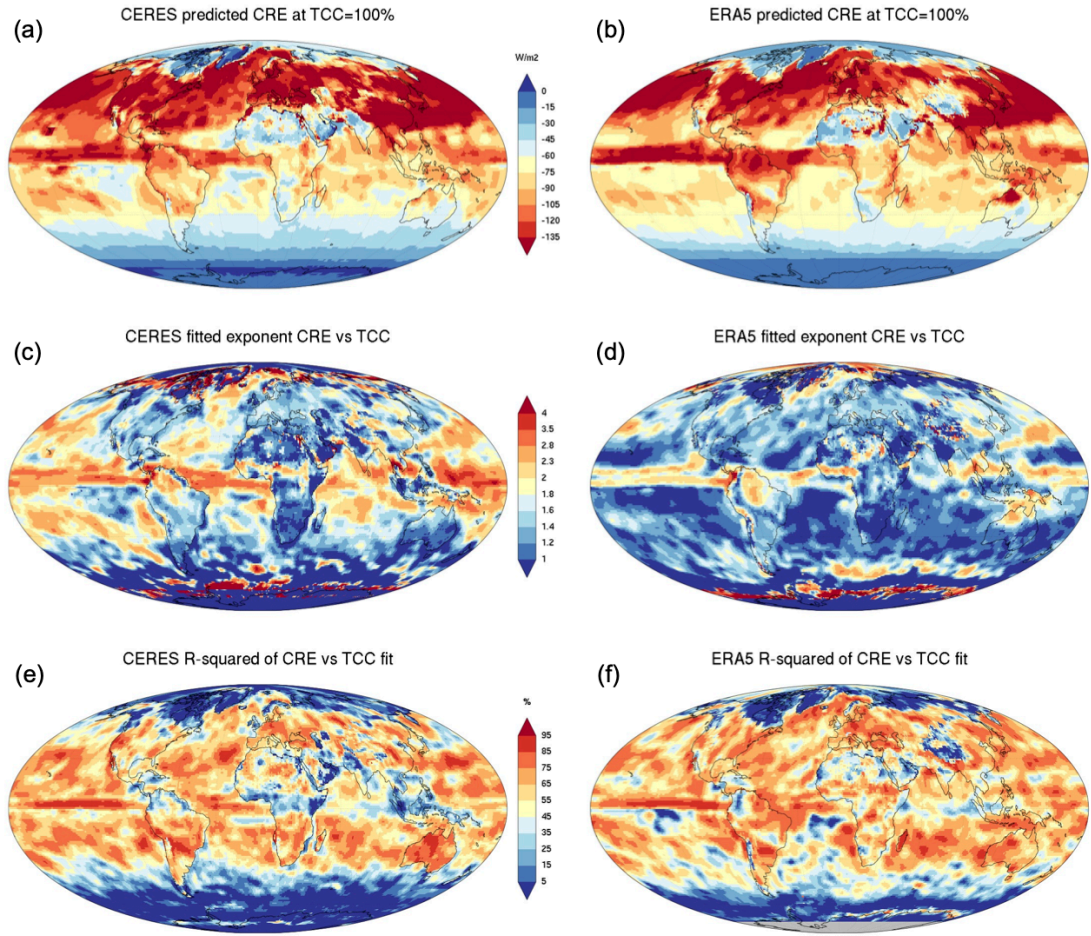

***Fig. S9. Parameters of the power functions fitted to the TCC-CRE-relation for May.*** *Based on inter-annual variations of monthly data 2001–2014, with May chosen arbitrarily as an example. (a) linear coefficient α corresponding to the predicted CRE at TCC=100% based on CERES; (b) same but based on ERA5; (c) exponent β with values above 1 indicating a superlinear dependence of CRE on TCC based on CERES; (d) same but based on ERA5; (e) Fraction of CRE variance explained by the fit based on CERES; (f) same but based on ERA5.*

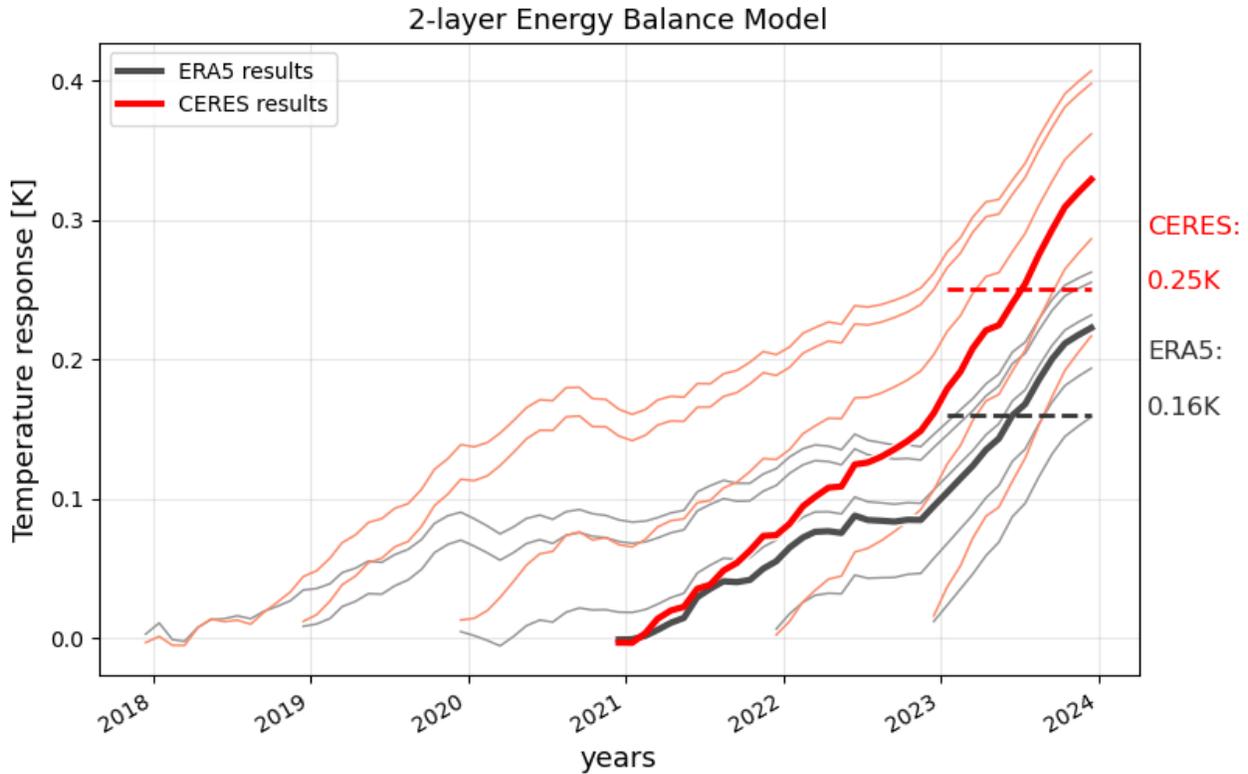

*Fig. S10. Upper-layer temperature response [K] to anomalous absorbed solar radiation from the CERES (red lines) and ERA5 (gray lines) datasets with respect to their 2001-2022 climatology.* The temperature estimate is computed with the 2-layer Energy Balance Model. Individual lines are initialized from December 2017 until December 2023, and then the solution to the EBM is found until December 2023. Horizontal dashed lines give the mean temperature response for the whole year 2023 (CERES: 0.25K, in red; ERA5: 0.16K, in black) for solutions initialized in December 2020, respectively.

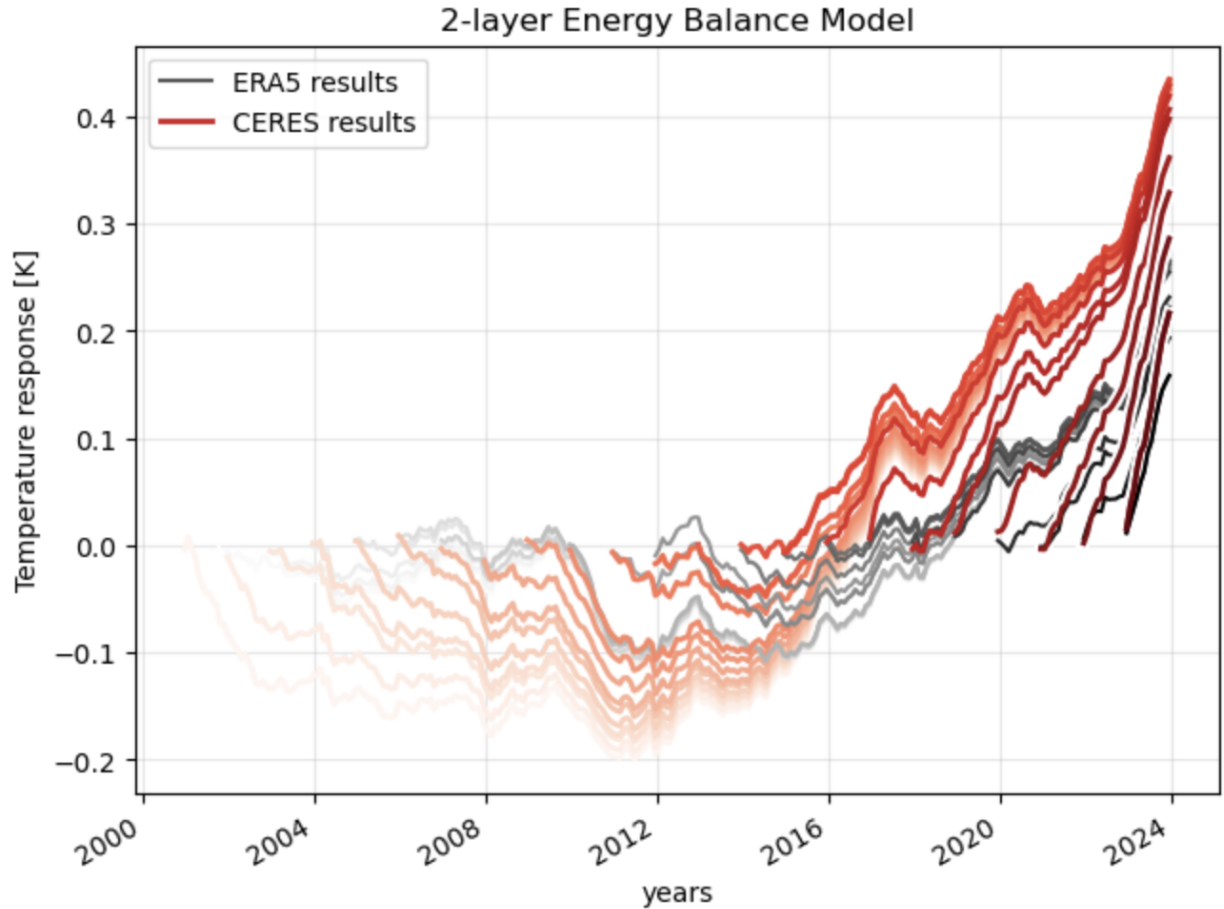

*Fig. S11. Upper-layer temperature response [K] to anomalous absorbed solar radiation from the CERES (red lines) and ERA5 (gray lines) datasets with respect to their 2001-2022 climatology.* The temperature estimate is computed with the 2-layer Energy Balance Model. Individual lines are initialized from all Decembers in the period December 2000 to December 2023, and then the solution to the EBM is found until December 2023.

*Table S1.*

*Parameters used in the 2-layer Energy Balance Model in this study. The parameters follow Hansen et al. (2005) and Geoffroy et al. (2013). By excluding the INM-CM4 model, the fitted ensemble-mean deep-ocean heat capacity $C_0$ has a much smaller standard deviation of 27 W yr m$^{-2}$ K$^{-1}$ (Geoffroy et al. 2013), indicating that the INM-CM4 might be an outlier in the multi-model CMIP ensemble, and therefore the parameters based on the remaining models are used here.*

| $F_{eff}$ (forcing efficacy) | $\lambda$ (radiative feedback parameter) | $\gamma$ (heat exchange coefficient) | $C$ (upper heat capacity) | $C_0$ (deep-ocean heat capacity) |
|---|---|---|---|---|
| 0.92 (Hansen et al., 2005) | 1.13 W m$^{-2}$ K$^{-1}$ (MMM in Table 3 in Geoffroy et al.) | 0.74 W m$^{-2}$ K$^{-1}$ (MMM in Table 4 in Geoffroy et al.) | 7.3 W yr m$^{-2}$ K$^{-1}$ (MMM in Table 4 in Geoffroy et al.) | 91.0 W yr m$^{-2}$ K$^{-1}$ (MMM in Table 4 in Geoffroy et al.) |